\documentclass[onecolumn,showpacs,preprintnumbers,amsmath,amssymb,11pt]{revtex4}

\usepackage{epsf}
\usepackage{graphicx}  % Include figure files
\usepackage{dcolumn}   % Align table columns on decimal point
\usepackage{bm}        % bold math
\usepackage{subfigure}

\newcommand{\be}{\begin{equation}}
\newcommand{\en}{\end{equation}}
\newcommand{\bea}{\begin{eqnarray}}
\newcommand{\ena}{\end{eqnarray}}
\begin{document}

\preprint{}

\title{Generalized Semi-Holographic Universe}

 \author{Hui Li \footnote{Electronic address: lihui@ytu.edu.cn} }
 \affiliation{
 Department of Physics, Yantai University, 30 Qingquan Road, Yantai 264005, Shandong Province, P.R.China }
\author{Hongsheng Zhang\footnote{Electronic address: hongsheng@shnu.edu.cn} }
\affiliation{Shanghai United Center for Astrophysics (SUCA),
 Shanghai Normal University, 100 Guilin Road, Shanghai 200234,
 P.R.China}
\author{Yi Zhang\footnote{Electronic address: zhangyia@cqupt.edu.cn}}
\affiliation{College of Mathematics and Physics, Chongqing University of
Posts and Telecommunications, Chongqing 400065, P.R.China}

\begin{abstract}
   We study the semi-holographic idea in context of decaying dark components. The energy flow between dark energy and the compensating dark matter is thermodynamically generalized to involve a particle number variable dark component with non-zero chemical potential. It's found that, unlike the original semi-holographic model, no cosmological constant is needed for a dynamical evolution of the universe. A transient phantom phase appears while a non-trivial dark energy-dark matter scaling solution keeps at late time, which evades the big-rip and helps to resolve the coincidence problem. For reasonable parameters, the deceleration parameter is well consistent with current observations. The original semi-holographic model is extended and it also suggests that the concordance model may be reconstructed from the semi-holographic idea.

\end{abstract}

\pacs{ 98.80.-k 95.36.+x 11.10.Lm} \keywords{holographic principle,
dark energy, grand canonical ensemble, chemical potential}

\maketitle

\section{Introduction}

 The existence of dark energy is one of the most significant cosmological discoveries
 over the last century \cite{acce}.  Various models of dark energy have been
proposed \cite{synthesis}, such as a small positive
 cosmological constant, quintessence, k-essence, phantom, quintom, etc., see Ref.\cite{review} for a
 recent review. Stimulated by the holographic principle, it's conjectured that dark energy problem may be a problem of quantum gravity, although its characteristic energy scale is very low\cite{cohn}. According to the holographic principle\cite{thsu}, the true laws of inside any surface are actually a description of  how its image evolves on that surface. In the case of black hole, the event horizon is a proper holography. When applying the holographic principle to the universe, the event horizon is probably not a good candidate of holography screen at least for two reasons: the identification of the cosmological event horizon is causally problematic in that it depends on the future of the cosmic evolution; a decelerating universe would have no holographic description, since it has no event horizon at all! With reference to the seminal work of Jacobson\cite{jaco}, several arguments have been put forward that the apparent horizon should be a causal horizon and is associated with the gravitational entropy and Hawking temperature\cite{cai} \cite{bak}, and hence the right holography of the universe. Apart from this key observation, other thermodynamical properties of the dark energy and dark matter have yet to be found in details\cite{entropyu}. Recently, the semi-holographic dark energy model\cite{semi-holo} has been proposed to examine how the universe could evolve if a dark component of the universe strictly obeys the holographic principle. It was found that, based on the first law of thermodynamics, the existence of the other dark component (dark matter) is compulsory, as a compensation of dark energy. In that model, with a non-zero cosmological constant the standard cosmological kinematics is recovered and the cosmological coincidence problem is also alleviated with the existence of a stable Einstein-de Sitter scaling solution.

   For all that, the properties of the semi-holographic model should be clarified for further estimations from three points of view. First, only massless particles (photons) exchange are considered. Or we treat dark matter and dark energy as two closed systems. Surely, by nature they can be open systems, and thus the particle numbers of the two systems can be variables. Second, the EOS of dark energy is constrained to be $\omega_{de}<-1$. It certainly is in accord with the latest cosmological observations and phenomenologically reasonable. However, as a thermodynamics motivated dark energy model, the construction would be founded on a more solid basis of thermodynamics and statistical physics, and exactly in this sense, the well-known theoretical difficulties does matter. As a matter of fact, the case of phantom energy with $\omega<-1$ is ruled out because the total entropy of the dark component is negative\cite{Lima2}. The equation of state (hereafter EOS) was restricted to the interval $-1\leq \omega<-1/2$ and a fermionic nature to the dark energy particles is favored. Thermodynamics arguments\cite{Gonzalez} in favor of the phantom hypothesis had to resort to an unusual assumption that the temperature of a phantomlike fluid is always negative in order to keep its entropy positive definite (as statistically required) or by arguing that the scalar field representations of a phantom field has a negative kinetic term which quantifies the translational kinetic energy of the associated fluid system. Third, as is known, the possibility of a coupling between dark matter and dark energy is often considered with three types in the literature, and those interaction forms are always introduced phenomenologically or through the low energy effective action\cite{chaplygin}: dark matter decaying into dark energy\cite{DM2DE}, dark energy decaying into dark matter\cite{DE2DM} and interaction in both directions\cite{DM-DE}. It's worth noting that, due to the non-adiabaticity of the holographic evolution of dark energy, the semi-holographic model obviously introduced a dark matter/dark energy non-gravitational interaction (A hypothetical dark energy property sometimes induces a non-gravitational coupling between dark matter and dark energy. See \cite{HuiLi} for another interacting model of this kind.), and the interaction belongs to the last category: dark matter dominated in the past and transferred energy to dark energy; the latter begins to dominate at present, and in the near future, the transfer would be reversed and dark energy decays to dark matter with a stable scaling solution reached. The theoretical obstacle is that, provided that the chemical potential of both components vanish, it is the decay of dark energy into dark matter that was favored by the thermodynamical point of view and the second law.

    Owing to a series of systematic work\cite{Lima2}\cite{Lima}\cite{Gonzalez}\cite{negtivemu}\cite{negtivemu2}, thermodynamics and statistical arguments on dark components and the cosmic evolution have come to several important conclusions which may kick off those dilemma. By considering the existence of a non-zero chemical potential, reanalysis of the thermodynamics and statistical properties of the dark energy scenario supports that the temperature of dark energy fluids must be always positive definite and that a negative chemical potential will recover the phantom scenario without the need to appeal to negative temperatures\cite{negtivemu}. In addition, a bosonic nature of the dark energy component becomes possible. If the chemical potential of at least one of the cosmic fluids is not zero, the decay can occur from the dark matter to dark energy, without violation of the second law\cite{negtivemu2}. From this point of view, the decay process between dark components with non-zero chemical potential is not merely a theoretical tool, but should be seriously considered as a prerequisite to turn the phantom dark energy and the decay in both directions to be physically real hypotheses.

          More importantly, due to the insufficiency in discussing the particle number variable transferring process between dark components, the original semi-holographic model should not be regarded as a complete thermodynamical examination of the semi-holographic idea in cosmology. A general framework dealing with the decay of the dark component particles is still in lack and therefore well motivated. Particularly speaking, the first law of thermodynamics would be applied in the context of grand canonical ensemble rather than canonical ensemble describing the original semi-holographic model.

    This paper is organized as follows: In the next section we will study the semi-holographic model in grand canonical ensemble with the dark energy chemical potential explicitly given. The dynamical analysis is left in section III. We find that there exists a stable dark matter-dark energy scaling solution at late time, which is helpful to address the coincidence problem. After a close analysis and comparison between model behavior and the observational facts, we present our conclusion and some discussions in section IV.

  \section{the model }
   There is decisive evidence that our observable universe evolves adiabatically
   after inflation in a comoving volume, that is, there is no energy-momentum
   flow between different patches of the observable universe so that the universe keeps
   homogeneous and isotropic after inflation.  That is the reason why we can
   use an FRW geometry to describe the evolution of the universe. In
   an  adiabatically evolving universe, the first law of
   thermodynamics equals the continuity equation. In
    a comoving volume the first law reads,
   \be
   dU=TdS-pdV+\tilde{\mu} dN,
   \label{1st}
   \en
   where $U=\Omega_k\rho a^3$ is the energy in this volume, $T$ denotes temperature,
   $S$ represents the entropy of this volume, $V$ stands for the physical
   volume $V=\Omega_k a^3$, $\tilde{\mu}$ is the chemical potential of the energy component with particle number $N$. Here, $\Omega_k$ is a factor related to the spatial curvature. For spatially flat case $\Omega_0=\frac{4}{3}\pi$, in this paper we only consider the spatially flat model, $\rho$ is the energy density and $a$ denotes the scale factor. The last term in the above equation indicates that the grand canonical ensemble instead of the canonical ensemble for the original semi-holographic model is considered. As a consequence, the framework allows decaying of the particles in consideration.

   The semi-holographic model concerns the possibility of a non-adiabatical dark energy, where the term $TdS$
   does not equal zero. Based on the investigations in \cite{bak, cai}, the entropy in
   a comoving volume the entropy becomes,
   \be
   S_c=\frac{8\pi^2 \mu^2}{H^2} \frac{a^3}{H^{-3}}={8\pi^2
   \mu^2}Ha^3.
   \label{Sc}
   \en
   where $H$ is the Hubble parameter, $\mu$ denotes the reduced
   Planck mass. The entropy has been reasonably assumed to be homogeneous in the observable universe. As our observable universe evolves adiabatically after reheating, the varying entropy of dark energy in a comoving volume should be compensated by an entropy change of the other component to keep the total entropy constant in a comoving volume.

    With the above supposition and conventions the entropy of the dark energy satisfies (\ref{Sc}),
    \be
    S_{de}={8\pi^2
   \mu^2}Ha^3.
   \label{detro}
    \en
    Correspondingly, the entropy of dark matter in this comoving
    volume should be
    \be
    S_{dm}=C-S_{de},
    \en
   where $C$ is a constant, representing the total entropy of the
   comoving volume.

   The Friedmann equation will determine the evolution of our universe
   \be
   H^2=\frac{1}{3\mu^2}(\rho_{dm}+\rho_{de}+\Lambda),
   \label{fried}
   \en
   where  H is the Hubble parameter, $\rho_{dm}$ denotes the density of non-baryon dark matter,
   $\rho_{de}$ denotes the density of dark energy, and $\Lambda$ is
   the cosmological constant (or vacuum energy). Partly because the partition of baryon matter is very
   small and does little work in the late time universe, and partly because the non-gravitational coupling between baryonic matter and dark energy is highly constrained, we just omit the baryon matter component in the discussion. The appearance of the cosmological constant in the expression is just for a general discussion, and as can be seen below, it is not a necessary  component in the generalized semi-holographic model. As a comparison, the cosmological constant has to be present in the original model, or else the ratio of holographic dark energy and dark matter density would always remain constant which is obviously not realistic.

   Below we will assume the chemical potential of dark energy to be of the form in Ref.\cite{negtivemu2}
   \be
   \tilde{\mu}=-\tilde\alpha T
   \en
   with the efficient $\tilde\alpha$ a positive constant. As was shown by Pereira, if the coefficient $\tilde\alpha>0$, a negative chemical potential will make the phantom fluid thermodynamically consistent and a decay of dark matter into dark energy possible. This assumption may help us highly broaden the exploration of the cosmological parameter region motivated by thermodynamics.

  \section{dynamical analysis}
  To investigate the evolution in a more detailed way, we take a
  dynamical analysis of the universe.

   The holographic principle requires that the temperature
   \cite{cai}
   \be
   T=\frac{H}{2\pi}.
   \label{tem}
   \en
   By using (\ref{tem}), (\ref{fried}), (\ref{detro}), and using the particle number $N$ to be proportional to the energy density, the first law of thermodynamics (\ref{1st}) becomes the evolution equation of dark energy,
   \be
   \frac{2}{3}\rho_{de} '=\frac{\rho_{dm}(1-w_{dm})-\rho_{de}(1+2 \tilde\alpha Y +3w_{de})+2\Lambda}{1+\tilde\alpha Y},
   \label{evlde}
   \en
    where a prime denotes the derivative with respect to $\ln a$, $w_{dm}$ indicates the EOS of dark matter, $w_{de}$ represents the EOS of dark energy and $Y=\sqrt{\rho_{de}+\rho_{dm}+\Lambda}$.
    Similarly, we derive the evolution equation of dark matter,

   \be
   \frac{2}{3}\rho_{dm} '=\frac{\rho_{de}[-1+(1-2\tilde\alpha Y)w_{de}]-\rho_{dm}[3+2\tilde\alpha Y+(1+2\tilde\alpha Y)w_{dm}]-2\Lambda}{1+\tilde\alpha Y}.
   \label{evldm}
   \en

   Clearly, the above equations will degenerate to the equations (8) and (9) of the paper \cite{semi-holo} when $\tilde\alpha=0$. For convenience we introduce two new dimensionless functions to represent the densities,
   \begin{eqnarray}
   &&u\triangleq \frac{\rho_{dm}}{3\mu^2H_0^2},\\
   &&v\triangleq \frac{\rho_{de}}{3\mu^2H_0^2},
   \end{eqnarray}
  and a dimensionless cosmological constant
  \be
  \lambda\triangleq \frac{\Lambda}{3\mu^2H_0^2},
  \en
  where $H_0$ denotes the present Hubble parameter. Then the
  equation set (\ref{evldm}), (\ref{evlde}) becomes
  \begin{eqnarray}
  &&\frac{2}{3} u '=\frac{-u[3+2\sqrt{3}\alpha y+(1+2\sqrt{3}\alpha y w_{dm})]+v[-1+(1-2\sqrt{3}\alpha y)w_{de}]-2\lambda}{1+\sqrt{3}\alpha y},
   \label{evldm1}\\
  && \frac{2}{3}v '=\frac{u(1-w_{dm})-v(1+2\sqrt{3}\alpha y+3w_{de})+2\lambda}{1+\sqrt{3}\alpha y}
   \label{evlde1}
   \end{eqnarray}
   respectively, where $y=\sqrt{u+v+\lambda}$ and $\alpha=\tilde\alpha\mu H_0$ are both dimensionless. We note that the  time variable does not appear in the dynamical system
  (\ref{evldm1}) and (\ref{evlde1}) because time has been completely replaced by scale factor.

  Before presenting the numerical examples for special parameters we
  study the analytical property of this system. And we will below take $\Lambda=0$ for simplicity.  The critical points of the
  dynamical system (\ref{evldm1}) and (\ref{evlde1}) are given by
  \be
  u_c'=v_c'=0,
  \en
  which yields,
  \be
  u_c=0,~~~~ v_c=0
  \label{solution1}
  \en
  and
  \be
  u_c=\frac{(1+w_{de})[1+w_{de}(2+w_{dm})]^2}{\tilde{\alpha}^2(w_{de}-w_{dm})(1+w_{dm})^2},~~~~v_c=-\frac{[1+w_{de}(2+w_{dm})]^2}{\tilde{\alpha}^2(w_{de}-w_{dm})(1+w_{dm})}.
  \label{solution2}
  \en

   The trivial solution corresponds to a future infinitely diluted universe and is less interesting for the coincidence problem. On the other hand, the non-trivial fixed point satisfies the below equality:
   \be
  \frac{u_c}{v_c}=-\frac{1+w_{\rm de}}{1+w_{\rm dm}},
  \en
  which is shared by the seminal model without chemical potential involved.
  So, finally the universe enters a de Sitter phase, and the ratio of dark matter over dark energy is independent of the chemical potential. Furthermore, there are two reasonable cases for the non-trivial scaling solution: case I, $w_{\rm de}<-1$ and
   $w_{\rm dm}>-1$; case II, $w_{\rm de}>-1$ and $w_{\rm dm}<-1$, since we should require that the final densities of dark matter and dark energy are both positive; as a matter of fact, the negative energy density may appear at most as a transient phenomenon, and it can not be a physically stable and permanent state.

To study the stability property of the fixed points\cite{stability}, the evolution equations should be perturbed to the first order and we list them in the Appendix.

 Then, the perturbation equations to the trivial critical point of \ref{solution1} are simplified as
 \begin{eqnarray}
 && \frac{2}{3}(\delta \rho_{dm})'=-\delta \rho_{dm}(3+w_{\rm dm})+\delta
  \rho_{de} (-1+w_{\rm de}),\\
 && \frac{2}{3}(\delta \rho_{de})'=\delta \rho_{dm}(1-w_{\rm dm})-\delta
  \rho_{de} (1+3w_{\rm de}).
  \end{eqnarray}

  And near this fixed point,the eigenvalues of the linearized system read
  \begin{eqnarray}
 && l_1=\frac{1}{2}(-4-3w_{de}-w_{dm}-\sqrt{-8w_{de}+9w_{de}^2+8w_{dm}-10 w_{de} w_{dm}+w_{dm}^2}),\\
  &&l_2=\frac{1}{2}(-4-3w_{de}-w_{dm}+\sqrt{-8w_{de}+9w_{de}^2+8w_{dm}-10 w_{de} w_{dm}+w_{dm}^2}).
 \end{eqnarray}

Stability requires that all of real parts of the eigenvalues are less
  than zero. Combined this requirement with the observational fact of $w_{de}<-\frac{1}{3}$, we find
  \be
  -1<w_{dm}<1,~~~~-\frac{1}{2+w_{dm}}<w_{de}<w_{dm}.
  \en

 \begin{figure}
 \centering
 \includegraphics[totalheight=2.3in, angle=0]{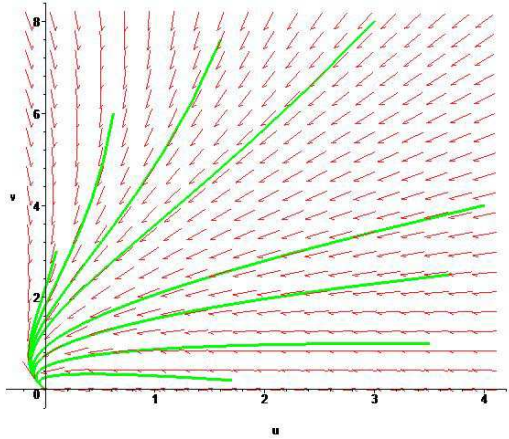}
 \caption{The plane v versus u. The initial conditions are taken for different orbits. There is a stationary node at the origin of coordinates, which attracts orbits in the first quadrant; however the orbits have to get through the v-axis before reaching the node, and therefore the case may be unrealistic.}
 \label{dece}
 \end{figure}

 A typical picture of the evolution is displayed in Figure 1. For illustration we set $w_{de}=-0.4$, $w_{dm}=-0.1$ and $\alpha=0.9$. As is seen to us, the orbits with initial values of energy densities lying in the first quadrant evolve to the second quadrant with the energy density of dark matter negative, and they will keep negative before the tracks reach the trivial critical point at the origin of coordinates. Evidently, this case is physically problematic and might be irrelevant to our universe.

 Now let's turn to the non-trivial fixed point. Analytic eigenvalues of the corresponding linearized system to this critical point have not been found, and we will explore the stability property of the scaling solution via numerical method. Here we plot Figure 2 to display the properties of evolution of the universe controlled by the dynamical system. In the figure, we set $\lambda=0$, $w_{de}=-1.2$, $w_{dm}=-0.2$ and the positive coefficient $\alpha=0.6$ for illustration. The physically meaningful region is just the first quadrant and different initial value tracks evolve to the common non-trivial fixed point still in the first quadrant, which indicates the existence of the physically stable attractor.

\begin{figure}\centering
 \subfigure[]
 {\includegraphics[width=65mm,height=75mm,bb=18 23 350 400]{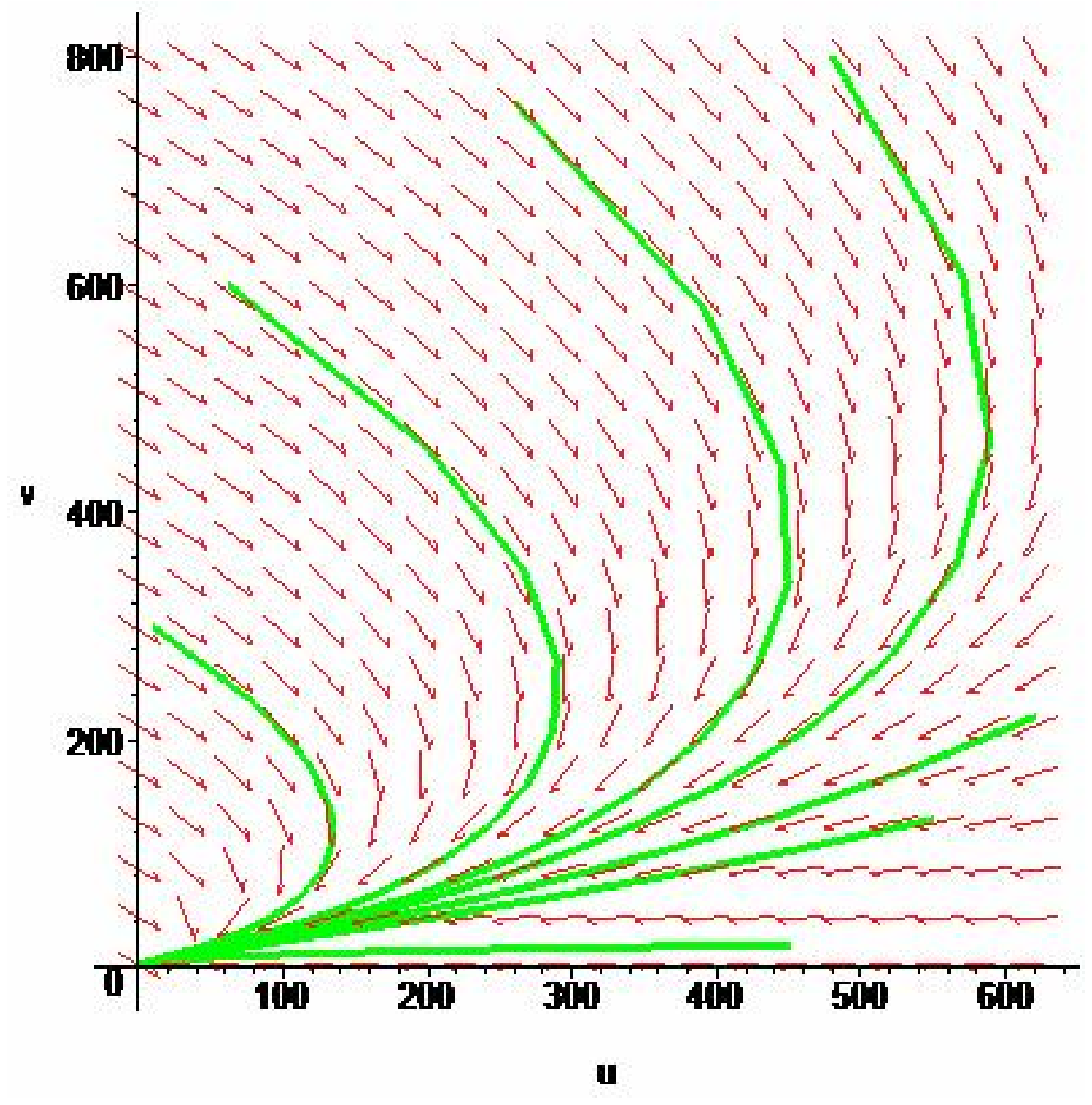}\label{1}}\quad
 \subfigure[]
 {\includegraphics[width=65mm,height=75mm,bb=18 23 350 400]{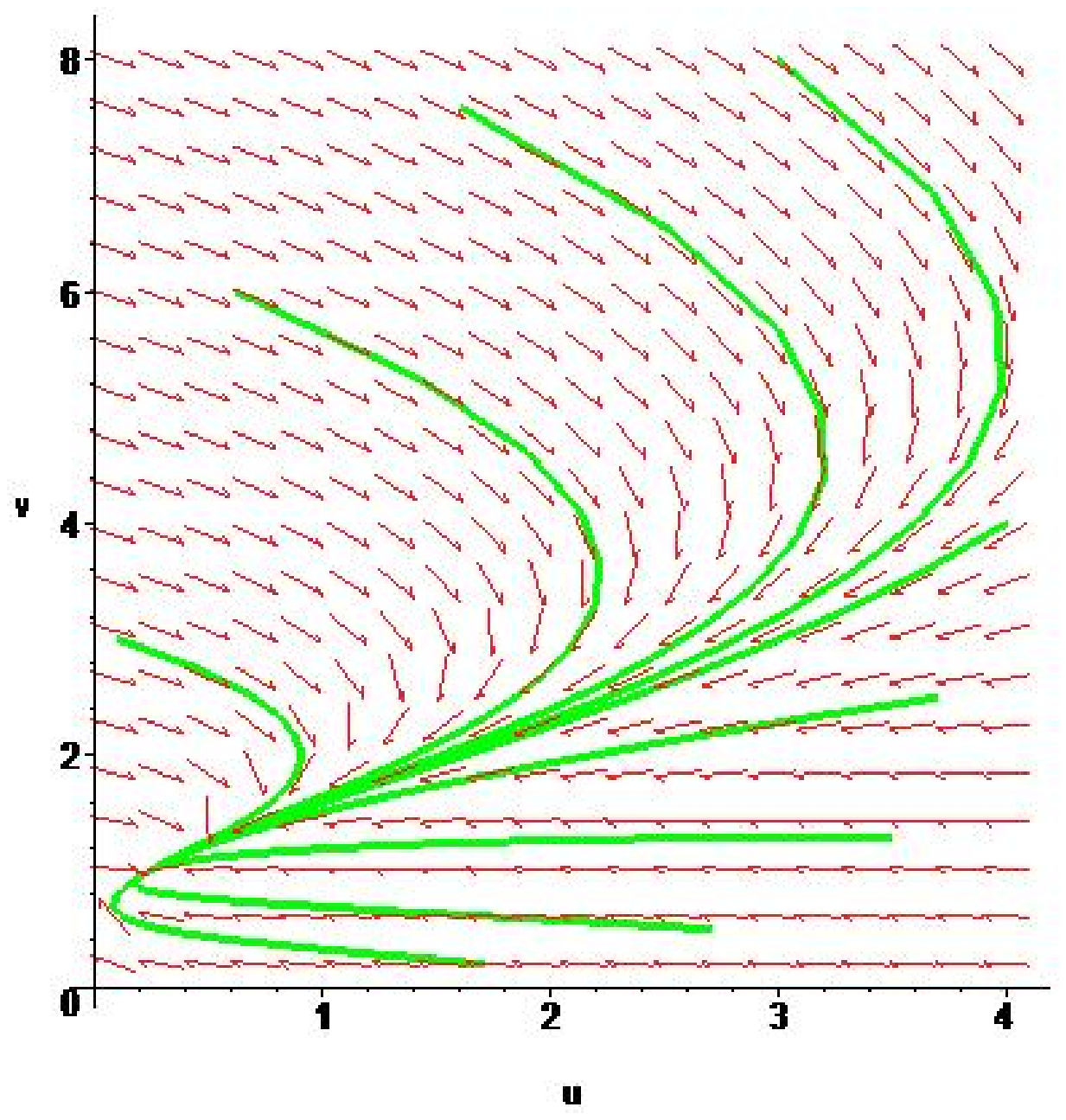}\label{2}}\quad
 \caption{The plane v versus u. (a)We consider the evolution of the universe. The initial conditions are taken for different orbits. It's clear that there is a stationary node, which attracts orbits in the first quadrant. (b)Orbits distributions around the node.}
 \label{attractor}
 \end{figure}

The most significant parameter from the viewpoint of
  observations is the deceleration parameter $q$, which carries the total
  effects of cosmic fluids. Using (\ref{fried}), (\ref{evlde}), and
  (\ref{evldm}) we obtain the deceleration parameter in this model
  \be
  q=-\frac{\ddot{a}}{a} \frac{1}{H^2}
  =\frac{1}{2}\frac{\rho_{dm}(1+3w_{dm})+\rho_{de}(1+3w_{de})-2\lambda}
  {\rho_{dm}+\rho_{de}+\lambda}.
  \en
For a numerical example, we take the present dark matter partition
  $u_0=0.3$ and the present dark energy partition
  $v_0=0.7$, which is favored by present observations\cite{acce}.

\begin{figure}
 \centering
 \includegraphics[totalheight=2.3in, angle=0]{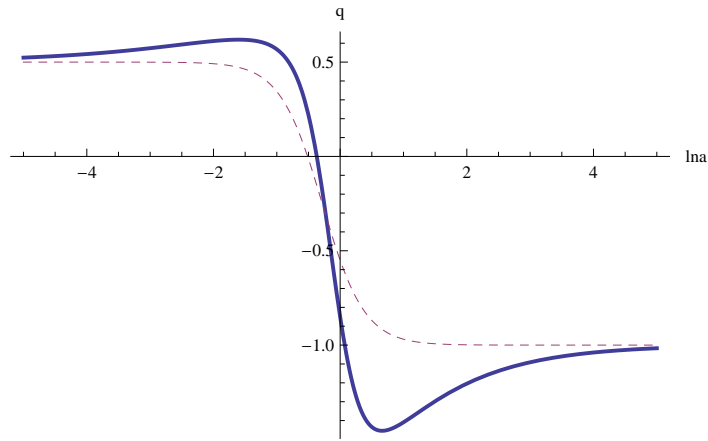}
 \caption{The evolutions of $q$ in the model (solid curve) and in $\Lambda$CDM (dashed curve), respectively.}
 \label{dece}
 \end{figure}

  Figure 3 illuminates the evolution of deceleration parameter. As a
  simple example we just set $w_{dm}=-0.2$, $w_{de}=-1.2$ and $\alpha=1.0$.
  The dark matter dominates in the past, and since its effective EOS in that period is larger than the assumed EOS, dark matter decays into dark energy. Only some time ago, this decay made the dark matter stiffer and with the accumulation of the dark energy, the universe endures a deceleration-acceleration transition. The acceleration goes through the present era and, if the dimensionless coefficient $\alpha$ is about $1$ or less, it will march for a phantom phase in the near future. That is to say, the deceleration parameter can be smaller than $-1$. However, the super-acceleration will not last for ever. The decay process of dark matter into dark energy then reverses and dark energy begins to transfer its energy to dark matter component. This conversion impedes the super-acceleration and eventually the evolution of the universe is pulled back to the Einstein-de Sitter track; therefore, the troublesome cosmic big-rip is evaded\cite{littlerip}. From the figure one sees that current $q\sim -1$ and at the high redshift region
  it goes to $0.5$, which is consistent with current observations \cite{review, WMAP}. As a comparison,
  the evolution of the deceleration parameter in a spatially flat $\Lambda$CDM is also plotted, in which the density parameter of dark matter $\Omega_{dm}=0.3$ too.

%Figure \figure{dece}.  as the  more swiftly approaches $0.5$ (standard dark matter model, SCDM for short) in this semi-holographic model than that of $\Lambda$CDM.

  The density evolution of the cosmic fluid does not depend on the
  assumed EOS above, but the effective EOS. We define the effective EOS
  as the following procedure. Supposing the dark matter evolves adiabatically itself,
  we obtain its evolution from (\ref{1st}),
  \be
 {d\rho_{dm}}+3(\rho_{dm}+p_{eff})\frac{da}{a}=0,
 \label{em}
 \en
 where $p_{eff}$ denotes the effective pressure of dark matter. Then
 we obtain
 \begin{eqnarray}
 &&w_{dme}\triangleq \frac{p_{eff}}{\rho_{dm}}=\frac{\frac{v}{u}[1-(1-2\alpha y)w_{de}]+2\frac{\lambda}{u}+3+2\alpha y +(1+2\alpha y)w_{dm}}{2+2\alpha y}-1,\\
 &&w_{dee}=-\frac{\frac{u}{v}(1-w_{dm})-(1+2\alpha y +3w_{de})+2\frac{\lambda}{v}}{2+2\alpha y}-1,
 \end{eqnarray}
 which is variable in the evolution history of the universe.
 Note that we have assumed both $w_{dm}$ and $w_{de}$ are constant from the beginning.
 Figure 4 displays the effective EOS of dark matter in which the same parameters are set as in Figure 3. It shows that, although its EOS is clearly negative, the dark matter behaves like cold dark matter until now. Only recently the dark matter has become stiff which relates to the current cosmic acceleration era. In Figure 5, different evaluations of the parameter $\alpha$ are taken and the corresponding deceleration evolution illustrated. A smaller $\alpha$ denotes a later deceleration-acceleration transition and a greater degree of acceleration in the near future. As the coefficient gets bigger, the cosmic evolution becomes closer to the concordant $\Lambda$CDM model and the transition happens more gently. For all that, the universe will go back to the same track of an everlasting de-Sitter phase with each of the energy densities unchanged thereafter.

   \begin{figure}
 \centering
 \includegraphics[totalheight=2.3in, angle=0]{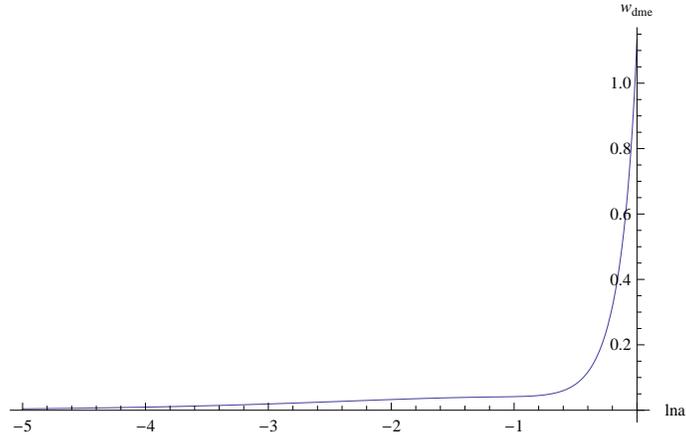}
 \caption{The effective EOS of dark matter $w_{dme}$ as a function of $\ln a$.}
 \label{weff11}
 \end{figure}.

   \begin{figure}
 \centering
 \includegraphics[totalheight=2.3in, angle=0]{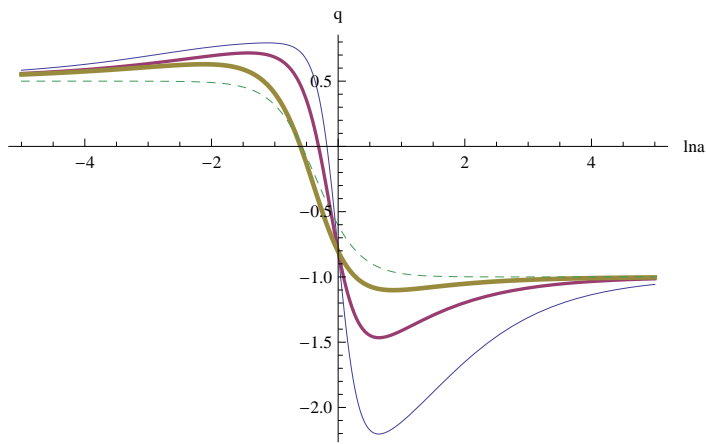}
 \caption{The evolutions of $q$ in semi-holographic model with different values of chemical potential coefficient $\alpha$ ($\alpha=0.1,0.6,1.2$ from thin to thick curves)and in $\Lambda$CDM (dashed curve), respectively.}
 \label{qmu}
 \end{figure}

The deceleration-acceleration transition happens abruptly in the original semi-holographic model, and the present model in grand canonical ensemble is prone to moderate the process. As the chemical potential coefficient changes, the deceleration parameter tracks fill in the intermediate region between the original semi-holographic model and the $\Lambda$CDM model. Therefore, in this sense we can say that the particle decay picture extends the original semi-holographic model moderately and reconstructs the concordance model with the semi-holographic idea. The tracks may have or may not have a super-acceleration phase which depends on the different values of $\alpha$.

%%%%%%%%%%%%%%%%%%%%%%%%%%%%%%%%%%%%%%%%%%%%%%%%%%%%%%%%%%%%%%%

\section{Conclusion and discussion}

 Based on the semi-holographic model inspired by holographic principle, especially the previous studies of the
 relation between thermal dynamics and general relativity, we find that, the semi-holographic dark energy model in the grand canonical ensemble identification recovers the standard expansion history and evolves to the Einstein-de Sitter final state through a possible transient super-acceleration.

 A phantom phase may appear in the near future without a big-rip in the end. Assuming the total matter
 in a comoving volume to be evolved adiabatically, a compensating dark matter component should exist and the non-gravitational coupling between dark energy and dark matter is also compulsory. As is shown in the above, this interaction yields a future attractor solution, which is a stable scaling solution for the dark matter-dark energy system in proper region of the
 parameters ($w_{\rm dm},~w_{\rm de}$). The final ratio of dark matter to dark energy
  only depends on $w_{\rm dm},~w_{\rm de}$, and is independent of the initial
 values of the densities of dark matter and dark energy and the coefficient $\alpha$ of chemical potential. This result
 is helpful to address the coincidence problem. The conversion of canonical ensemble context to grand canonical ensemble one seems to deform the original model to the well established $\Lambda$CDM model in the sense that different tracks roughly fill in the zone between the two models in the $q-\ln a$ plane. Therefore, the particle decay picture extends the original semi-holographic model moderately and reconstructs the concordance model with the semi-holographic idea. This grand canonical ensemble viewpoint opens the possibility to deepen our understandings of dark energy thermodynamics and statistical properties, and is closely related to previous systematic application of the cosmological thermodynamics.

  With numerical examples, it is illustrated that the deceleration parameter is well consistent with observations. As a phenomenological model, the parameters including $\alpha$ may be further constrained by observational data and detailed investigations on global fitting of cosmological parameters will also be interesting.

 {\bf Acknowledgments.}
 H. Li is supported by National Natural Science Foundation of China under grant No. 10747155, H. Zhang is supported the Program for Professor of Special Appointment (Eastern Scholar) at Shanghai Institutions of Higher Learning, Shanghai Municipal Pujiang grant No. 10PJ1408100, and National Natural Science Foundation of China under grant No. 11075106, and Y. Zhang is supported by
 the National Natural Science Foundation of China under grant Nos. 11005164, 11175270 and 10935013, the Distinguished Young Scholar Grant 10825313, CQ CSTC under grant No. 2010BB0408, and CQ MEC under grant No. KJTD201016.

 \section{The Appendix: The perturbation equations}
The perturbation equations for the combination $(\rho_{de},\rho_{dm})$ are:

    \begin{eqnarray}
      \nonumber
\delta \rho_{de}'= &&\frac{3}{2}\delta\rho_{dm}\left[\frac{1-w_{dm}-\frac{\tilde\alpha\rho_{de}}{Y}}{1+\tilde\alpha Y}+\frac{\rho_{de}(1+2\tilde\alpha Y+3w_{de})\frac{\tilde\alpha}{2 Y}-\rho_{dm}(1-w_{dm})\frac{\tilde\alpha}{2 Y}}{(1+\tilde\alpha Y)^2}\right]\\
 &&-\frac{3}{2}\delta\rho_{de} \left[\frac{\rho_{de}\frac{\tilde\alpha}{Y}+1+2\tilde\alpha Y+3w_{de}}{1+\tilde\alpha Y}+\frac{\rho_{dm}(1-w_{dm})\tilde\alpha-\rho_{de}(1+2\tilde\alpha Y+3w_{de})\tilde\alpha}{2Y(1+\tilde\alpha Y)^2}\right],
 \end{eqnarray}

\begin{eqnarray}
\nonumber
(\delta \rho_{dm})'=&&-\frac{3}{2}\delta\rho_{dm}\left[\frac{3+2\tilde\alpha Y+(1+2\tilde\alpha Y)w_{dm}+\rho_{dm}(1+w_{dm})\frac{\tilde\alpha}{Y}+\rho_{de}w_{de}\frac{\tilde\alpha}{Y}}{1+\tilde\alpha Y}\right.\\
\nonumber
 &&-\left.\frac{\rho_{de}[1-(1-2\tilde\alpha Y)w_{de}]\frac{\tilde\alpha}{2Y}+\rho_{dm}[3+2\tilde\alpha Y+(1+2\tilde\alpha Y)w_{dm}]\frac{\tilde\alpha}{2Y}}{(1+\tilde\alpha Y)^2}\right]\\
 \nonumber
  &&- \frac{3}{2}\delta\rho_{de}\left[\frac{\rho_{de}w_{de}\frac{\tilde\alpha}{Y}+\rho_{dm}(1+w_{dm})\frac{\tilde\alpha}{Y}+1-(1-2\tilde\alpha Y)w_{de}}{1+\tilde\alpha Y}\right.\\
   && -\left.\frac{\rho_{de}[1-(1-2\tilde\alpha Y)w_{de}]\frac{\tilde\alpha}{2Y}+\rho_{dm}[3+2\tilde\alpha Y+(1+2\tilde\alpha Y)w_{dm}]\frac{\tilde\alpha}{2Y}}{(1+\tilde\alpha Y)^2}\right]
\end{eqnarray}
with all the notations declared in the main text.

\end{document}